\documentclass[aps,prd,onecolumn,superscriptaddress,nofootinbib,11pt]{revtex4-2}

\usepackage{amsmath,amssymb,amsfonts,mathrsfs}
\usepackage{graphicx}
\usepackage{bm}
\usepackage{hyperref}
\usepackage{color}

\begin{document}

\title{Universality and Falsifiability of Quantum Spacetime Decoherence:\\
A Gauge-Invariant Framework for Gravitational-Wave Phase Diffusion}

\author{Hu Cang}
\email{cangh@uci.edu}
\affiliation{University of California, Irvine, Irvine, California 92697, USA}

\author{Yuan Wang}
\affiliation{The University of Hong Kong, Pokfulam Road, Hong Kong, China}


\begin{abstract}
We develop a fully gauge-invariant and rigorously derived framework for computing the cumulative decoherence of gravitational waves (GWs) propagating through a stochastic quantum spacetime. Working directly with the Riemann-tensor two-point function and exploiting the extreme adiabaticity of cosmological GW propagation ($H/\omega \ll 10^{-20}$), we show that phase diffusion—rather than amplitude attenuation or mode mixing—is the unique leading-order imprint of microscopic curvature fluctuations.

Our main theoretical result is a universality theorem: for \emph{any} quantum-gravity model whose curvature fluctuations possess a finite correlation length $L_c \ll D$ (with $D$ the GW propagation distance), the accumulated phase variance grows linearly with distance, $\beta = 1$, independent of the underlying microphysics. This diffusive scaling contrasts sharply with coherent astrophysical effects and with nonlocal models predicting $\beta \neq 1$. The frequency exponent $\alpha$ therefore becomes a clean spectral discriminator, separating string-foam recoil ($\alpha \simeq 4$), holographic or scale-invariant noise ($\alpha \simeq 2$), and causal-set discreteness ($\alpha \simeq 0$).

We obtain these results from first principles by evaluating the projected Riemann correlator along null geodesics and determining the exact conditions under which deviations from universality can arise. Finally, we outline a hierarchical Bayesian strategy for measuring $(A,\alpha,\beta)$ with LIGO/Virgo/KAGRA, LISA, and Pulsar Timing Arrays. Although standard Planck-scale fluctuations remain far below current sensitivity, this framework provides a sharp and falsifiable test of exotic quantum-spacetime scenarios, particularly those with macroscopic correlation lengths or strong energy dependence.
\end{abstract}

\maketitle

\section{Introduction}
The dawn of gravitational-wave (GW) astronomy has transformed the search for
quantum-gravity signatures. 
Since LIGO's first detection in 2015~\cite{LIGO2016}, gravitational-wave
astronomy has entered a precision era. The most recent LVK event catalog,
GWTC-3~\cite{GWTC3}, contains hundreds of compact-binary mergers spanning
frequencies from $\sim 10~\mathrm{Hz}$ to a few kHz and luminosity distances of
several gigaparsecs. Planned space-based detectors such as
LISA~\cite{LISA2017}, together with Pulsar Timing Arrays (PTAs), which have
recently reported evidence for a nanohertz gravitational-wave background
\cite{NANOGrav2023}, will extend this reach to millihertz and nanohertz
bands, enabling truly multi-band GW spectroscopy of the Universe.
Across these platforms, phase-coherent signals with
signal-to-noise ratios ${\rm SNR}\gtrsim 50$ offer sub-radian phase resolution
and exquisite sensitivity to tiny propagation effects accumulated over
cosmological baselines.

Most existing tests of quantum gravity in the GW sector have focused on local
corrections to classical wave propagation, such as Lorentz-violating dispersion
relations or birefringence, often constrained by multimessenger observations and
parameterized post-Einsteinian frameworks; see, e.g.,
Refs.~\cite{AmelinoCamelia1999,FordHu2008} and references therein. These tests
primarily probe corrections to the \emph{local} dispersion relation and are
largely insensitive to intrinsically nonlocal features of quantum spacetime,
such as stochastic curvature fluctuations on Planckian or mesoscopic scales.

\begin{figure}[t]
    \centering
    \includegraphics[width=0.9\linewidth]{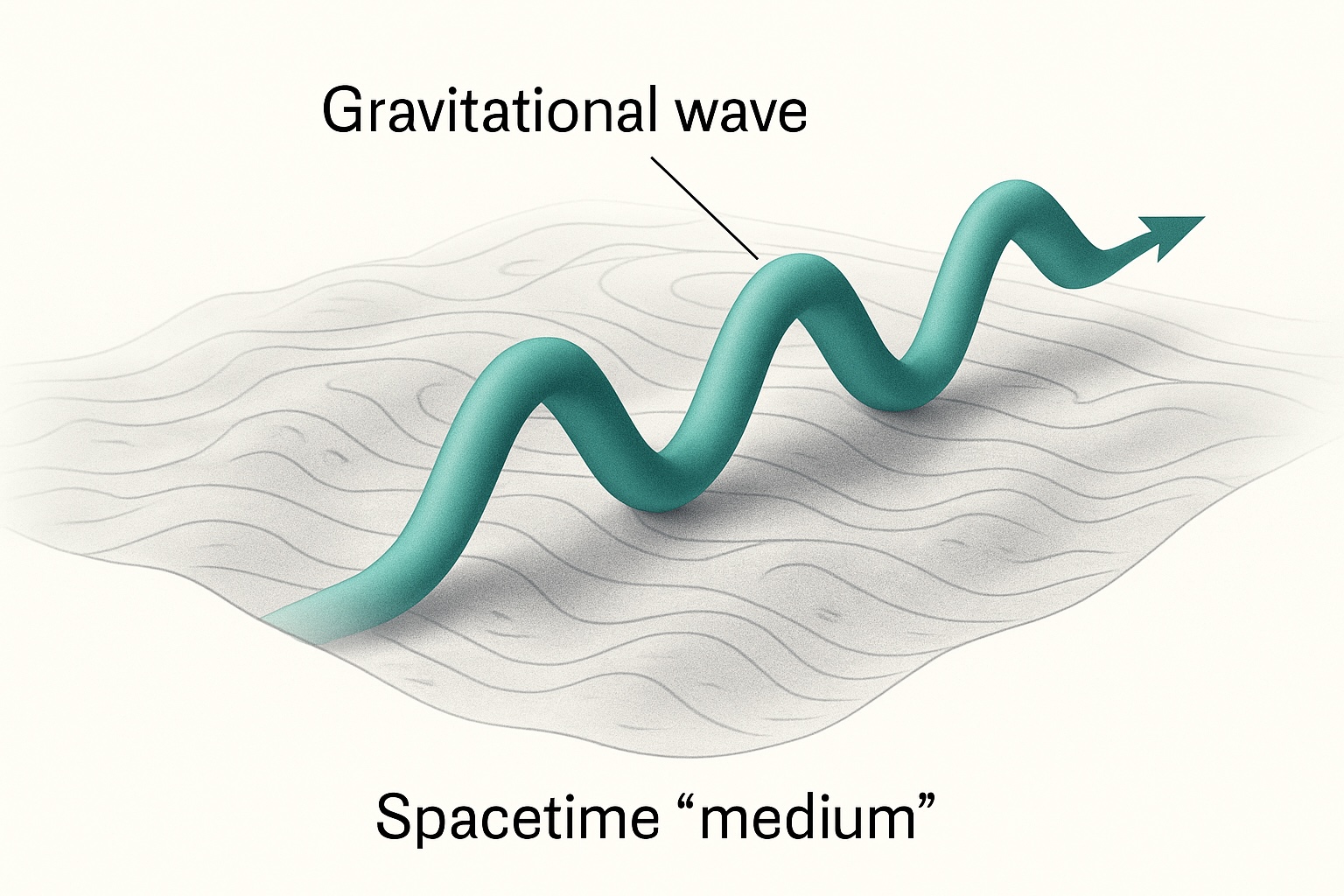}
    \caption{Schematic illustration of gravitational-wave propagation through a fluctuating spacetime background. A coherent gravitational wave (blue) samples microscopic fluctuations of the geometry (gray surface) along a cosmological trajectory. In the adiabatic regime, phase coherence is preserved and microscopic fluctuations contribute primarily through cumulative phase shifts.}
    \label{fig:spacetime_fluctuations}
\end{figure}

In this work, we instead exploit a qualitatively different observational window:
the ability to track the phase evolution of a coherent wave over gigaparsec
baselines. This vast lever arm allows one to probe the \emph{cumulative}
influence of microscopic spacetime fluctuations---a fundamentally nonlocal
effect---which cannot be accessed in terrestrial or solar-system experiments; a
schematic illustration is shown in Fig.~\ref{fig:spacetime_fluctuations}.

Formulated in this way, the central question becomes: how do microscopic
curvature fluctuations modify the phase of an adiabatically propagating GW, and
how do those modifications scale with frequency and propagation distance?
We address this question by constructing a manifestly gauge-invariant framework
based on the Riemann tensor two-point function projected along a null geodesic. 
To ensure physical meaningfulness, we depart from previous approaches that rely on
metric perturbations, which suffer from gauge ambiguities. Instead, we construct
a manifestly gauge-invariant theory based on the Riemann tensor two-point
function projected along a null geodesic. The physical observable is the GW
phase $\phi$, whose stochastic deviation $\Delta\phi$ accumulates continuously
along the trajectory (Fig. \ref{fig:spacetime_fluctuations}).

The primary contribution of this paper is a rigorous derivation of the scaling law
\begin{equation}
\label{eq:scaling}
\langle \Delta\phi^2 \rangle = A^2 \left(\frac{\omega}{\omega_0}\right)^{\alpha} \left(\frac{D}{D_0}\right)^{\beta}.
\end{equation}
We prove a universality theorem stating that for any model with short-range curvature correlations, the distance exponent saturates to $\beta=1$. This result transforms the search for quantum gravity from an ambiguous hunt for “noise” into a precise spectroscopic measurement of the exponents $(\alpha, \beta)$. We further demonstrate that while standard Planckian fluctuations are strongly suppressed, more exotic models with macroscopic correlation lengths or enhanced energy dependence are already amenable to constraint by current and future detectors.

\section{Adiabatic propagation: physical and mathematical foundations}

We first justify why phase diffusion dominates over all other potential propagation effects. Gravitational waves satisfy the geometric-optics (WKB) ansatz
\begin{equation}
h_{\mu\nu}(x) = A_{\mu\nu}(x)\,e^{i\phi(x)},
\end{equation}
where the phase varies on timescales of order $\omega^{-1}$, while the amplitude varies only on cosmological scales. The background curvature scale is set by the Hubble rate $H$ and by large-scale inhomogeneities. Any interaction between the GW and the background geometry that induces mode conversion, backscattering or dissipation necessarily involves curvature Fourier modes with characteristic wavenumbers comparable to the GW wavenumber $k \sim \omega$. In a cosmological background, however, the dominant curvature modes satisfy $k \sim H$, so the corresponding mixing amplitudes are suppressed by powers of $(H/\omega)^n$ with $n\ge 2$.

By contrast, the phase
\begin{equation}
\phi = \int k_\mu\,dx^\mu
\end{equation}
is not an adiabatic invariant. Tiny perturbations of the null wavevector $k^\mu$ induced by microscopic curvature fluctuations lead to small shifts in the instantaneous frequency $\omega$, but these shifts integrate coherently along the ray and can accumulate over cosmological distances. The net effect is a stochastic, diffusive phase shift whose variance grows with the path length $D$.

To formalize this, consider a comoving observer with four-velocity $u^\mu$ measuring a frequency
\begin{equation}
\omega = - k_\mu u^\mu.
\end{equation}
By differentiating along an affine parameter $\lambda$ and using the geodesic deviation equation, one can derive an exact, gauge-invariant expression for the perturbation to the measured frequency. Restricting attention to the electric components of the Riemann tensor that couple to the polarization tensor of the GW, one obtains
\begin{equation}
\label{eq:deltaomega}
\frac{d (\delta \omega)}{d\lambda}
=
-\frac{1}{2\omega}\,
R_{\mu\nu\rho\sigma}\,
k^\mu k^\rho \epsilon^\nu \epsilon^\sigma,
\end{equation}
where $\epsilon^\mu$ is a parallel-propagated polarization vector satisfying $k_\mu \epsilon^\mu=0$ and $\epsilon_\mu \epsilon^\mu=1$. Equation~(\ref{eq:deltaomega}) is manifestly gauge-invariant and encapsulates the physical tidal effect of curvature on the GW phase.

Integrating Eq.~(\ref{eq:deltaomega}) along the null geodesic, and assuming that the mean of the microscopic fluctuations vanishes, $\langle \delta\omega \rangle = 0$, the accumulated phase shift is
\begin{equation}
\Delta\phi = \int_0^D \delta\omega(s)\,ds,
\end{equation}
where $s$ is the proper distance (or an affine parameter proportional to it) along the ray. The ensemble-averaged phase variance then takes the form
\begin{equation}
\label{eq:phasevariance}
\langle \Delta\phi^2\rangle
= 
\frac{1}{4\omega^2}
\int_0^D \!ds \int_0^D \!ds'\,
C_R(s-s'),
\end{equation}
where the projected Riemann correlator $C_R(\Delta s)$ is defined by
\begin{equation}
C_R(\Delta s)
=
\big\langle
R_{\mu\nu\rho\sigma}(s)\,
R_{\alpha\beta\gamma\delta}(s')
\big\rangle
\,k^\mu k^\rho k^\alpha k^\gamma
\epsilon^\nu \epsilon^\sigma \epsilon^\beta \epsilon^\delta,
\qquad \Delta s = s-s'.
\end{equation}
Equation~(\ref{eq:phasevariance}) is the master formula of this work, connecting microscopic spacetime fluctuations to macroscopic decoherence observables in a fully covariant and gauge-invariant manner.

\section{Stochastic spacetime structure and assumptions}

To make progress with Eq.~(\ref{eq:phasevariance}) we must specify the statistical properties of the fluctuation field. We consider a random tensor field $\delta R_{\mu\nu\rho\sigma}(x)$ defined on a smooth background $g^{(0)}_{\mu\nu}$, with $\langle\delta R_{\mu\nu\rho\sigma}\rangle=0$, and we assume statistical homogeneity and stationarity along the GW trajectory. In particular, the projected correlator $C_R(s-s')$ is taken to depend only on the affine separation $\Delta s = |s-s'|$ along the ray, which is justified whenever the microscopic physics is approximately translation invariant on scales much smaller than the Hubble radius.

A key assumption is the existence of a finite correlation length $L_c$ such that $C_R(\Delta s)$ decays rapidly for $|\Delta s| \gg L_c$. For Planckian foam one typically expects $L_c\sim \ell_P$, while in models with large extra dimensions or nontrivial UV completions $L_c$ could be larger, up to $L_c\sim 1/\mathrm{TeV}$. If curvature correlations extended coherently over gigaparsecs, the spacetime would exhibit intrinsically nonlocal or fractal structure on macroscopic scales, which we treat separately as a distinct class of models.

The frequency dependence of the fluctuations is encoded in the variance of the projected curvature element
\begin{equation}
R_{\rm eff}(x)
\equiv
R_{\mu\nu\rho\sigma}(x)
k^\mu k^\rho \epsilon^\nu \epsilon^\sigma.
\end{equation}
Since $k^\mu\sim \omega$ for a GW of angular frequency $\omega$, one expects
\begin{equation}
\langle R_{\rm eff}^2\rangle \propto \omega^{4+p},
\end{equation}
where the exponent $p$ captures the microscopic energy dependence of the curvature fluctuations. Scale-invariant models such as Planckian white noise or holographic noise correspond to $p=0$, while energy-dependent scattering in noncritical string theory can yield $p>0$.
It is convenient to denote the microscopic variance of the projected curvature by
\begin{equation}
\sigma_R^2 \equiv \langle R_{\rm eff}^2\rangle,
\end{equation}
so that $\sigma_R$ sets the typical amplitude of curvature “kicks” experienced by
the GW. Together with the correlation length $L_c$ introduced above, $\sigma_R^2$
will later combine into an effective diffusion coefficient that controls the
macroscopic phase variance.

\section{Derivation of the scaling law and universality of $\beta=1$}

We now evaluate Eq.~(\ref{eq:phasevariance}) and derive the scaling exponents. Using the symmetry of the integrand, we can rewrite the double integral as
\begin{equation}
\langle \Delta\phi^2\rangle
=
\frac{1}{2\omega^2}
\int_0^D (D-u) C_R(u)\,du,
\label{eq:Du}
\end{equation}
where we have introduced $u = |s-s'|$ and exploited the identity
\begin{equation}
\int_0^D ds \int_0^D ds'\,f(|s-s'|)
=
2\int_0^D (D-u)f(u)\,du.
\end{equation}

In the regime of short-range correlations, the integral of the correlator converges
to a finite value, which can be expressed in terms of the microscopic variance
$\sigma_R^2$ and correlation length $L_c$ as
\begin{equation}
\sigma_R^2 L_c \equiv \int_0^\infty C_R(u)\,du < \infty,
\end{equation}
and $C_R(u)$ is negligible for $u\gtrsim L_c$. We can then write
\begin{equation}
\int_0^D (D-u) C_R(u)\,du
=
D\int_0^D C_R(u)\,du - \int_0^D u\,C_R(u)\,du.
\end{equation}
For $D\gg L_c$, both integrals saturate as $D\to\infty$:
\begin{equation}
\int_0^D C_R(u)\,du \to \sigma_R^2 L_c,
\qquad
\int_0^D u\,C_R(u)\,du \to \ell_1,
\end{equation}
where $\ell_1$ is a finite constant determined by the short-distance structure of the correlator. Thus,
\begin{equation}
\int_0^D (D-u) C_R(u)\,du
=
D\sigma_R^2 L_c - \ell_1 + \mathcal{O}(e^{-D/L_c}),
\end{equation}
and the subleading constant and exponentially small corrections do not affect the scaling with $D$.

Substituting into Eq.~(\ref{eq:Du}) yields
\begin{equation}
\label{eq:finalbeta1}
\langle \Delta\phi^2\rangle
=
\frac{1}{2\omega^2}
\left[
D\sigma_R^2 L_c - \ell_1 + \ldots
\right].
\end{equation}
For $D\gg L_c$ the dominant term is
\begin{equation}
\langle \Delta\phi^2\rangle \propto \frac{\sigma_R^2 L_c}{\omega^2}\,D.
\end{equation}
We thus obtain the universal distance scaling
\begin{equation}
\beta = 1.
\end{equation}

This is the precise content of our universality theorem: for any microscopic curvature correlator that is integrable (short-range) along the ray, the accumulated phase variance grows linearly with distance, reflecting a random walk in phase induced by a large number of independent curvature “kicks.” The universality is a geometric analogue of the Central Limit Theorem applied to spacetime fluctuations.

\section{Conditions for $\beta\neq 1$ and long-range correlations}

Deviations from $\beta=1$ require a breakdown of the short-range assumption. A natural parametrization of long-range correlations is a power-law tail,
\begin{equation}
C_R(u) \sim u^{-\gamma},\qquad 0 < \gamma < 1,
\end{equation}
for which the integral $\int_0^\infty C_R(u)\,du$ diverges. In this case the scaling of Eq.~(\ref{eq:Du}) changes qualitatively. One finds
\begin{equation}
\int_0^D (D-u) u^{-\gamma}\,du
=
D^{2-\gamma} \int_0^1 (1-x)x^{-\gamma}\,dx,
\end{equation}
so that
\begin{equation}
\langle \Delta\phi^2\rangle \propto \omega^{-2} D^{2-\gamma},
\end{equation}
and hence
\begin{equation}
\beta = 2 - \gamma,\qquad 1 < \beta < 2.
\end{equation}
This regime corresponds physically to nonlocal or fractal spacetime structure, in which curvature fluctuations remain correlated over macroscopic distances. Phenomenological examples include infrared-modified gravity with nonlocal propagators, spacetimes with effective fractional dimension, or highly coherent astrophysical media. An observational measurement of $\beta\neq 1$ would therefore indicate a striking breakdown of locality, whereas a confirmation of $\beta=1$ would rule out this entire class of nonlocal quantum-spacetime scenarios.

\section{Frequency scaling and the exponent $\alpha$}

The frequency exponent $\alpha$ arises from the explicit frequency dependence of the curvature variance. In the short-range regime, the dominant scaling is captured by
\begin{equation}
\langle \Delta\phi^2\rangle
\sim
\frac{1}{\omega^2}\,\langle R_{\rm eff}^2\rangle L_c\,D
\sim
\omega^{(4+p)-2} L_c D
=
\omega^{2+p} L_c D,
\end{equation}
and we define
\begin{equation}
\alpha = 2 + p,
\qquad
\beta = 1.
\end{equation}
Scale-invariant quantum foam or holographic models, where the microscopic curvature variance does not depend explicitly on $\omega$ beyond the kinematic contractions, correspond to $p=0$ and hence $\alpha=2$. Energy-dependent models such as D-brane recoil in string theory can yield $p>0$ and therefore larger values of $\alpha$.

\section{Worked example: stringy spacetime foam}

To illustrate how the microscopic physics maps onto $\alpha$, we consider a string-inspired D-brane recoil model. In such scenarios, a graviton of energy $E\sim\omega$ scatters off a fluctuating D-particle defect, inducing a local metric disturbance that depends on the energy of the probe,
\begin{equation}
\delta g_{\mu\nu} \sim \left(\frac{E}{M_P}\right)^n
\sim \left(\frac{\omega}{M_P}\right)^n,
\end{equation}
for some integer $n\ge 0$. Curvature involves two derivatives of the metric, so the fluctuations in the Riemann tensor scale as
\begin{equation}
\delta R_{\mu\nu\rho\sigma} \sim \partial^2 \delta g_{\mu\nu} \sim k^2 \delta g_{\mu\nu} \sim \omega^2 \left(\frac{\omega}{M_P}\right)^n.
\end{equation}
The projected curvature that enters the phase shift has two additional factors of $k^\mu$,
\begin{equation}
R_{\rm eff} \sim R_{\mu\nu\rho\sigma}k^\mu k^\rho \epsilon^\nu \epsilon^\sigma \sim \omega^2 \left(\frac{\omega}{M_P}\right)^n \times \omega^2 = \omega^{4+n} M_P^{-n}.
\end{equation}
The variance then scales as
\begin{equation}
\langle R_{\rm eff}^2\rangle \sim \omega^{8+2n} M_P^{-2n}.
\end{equation}
Substituting into the short-range expression for the phase variance,
\begin{equation}
\langle \Delta\phi^2\rangle \sim \frac{1}{\omega^2}\,\langle R_{\rm eff}^2\rangle L_c\,D
\sim \omega^{6+2n} M_P^{-2n} L_c D.
\end{equation}
This corresponds to
\begin{equation}
\alpha = 6+2n - 4 = 2 + 2n,
\end{equation}
where the subtraction of 4 removes the two explicit $\omega^2$ factors from the kinematic contractions and the $1/\omega^2$ prefactor. For the linear recoil case $n=1$ one obtains $\alpha=4$, while $n=0$ recovers $\alpha=2$, consistent with the scale-invariant limit. Thus, string-foam models naturally populate the region $\alpha\approx 4$--$6$ along the universal $\beta=1$ line.

\section{Normalization and physical meaning of $A$}

The phenomenological amplitude $A$ in Eq.~(\ref{eq:scaling}) encodes the overall strength of the curvature fluctuations. In the short-range regime, comparing Eq.~(\ref{eq:finalbeta1}) with the scaling form
\begin{equation}
\langle \Delta\phi^2\rangle = A^2 \left(\frac{\omega}{\omega_0}\right)^{\alpha} \left(\frac{D}{D_0}\right)^{\beta},
\end{equation}
with $\alpha=2+p$ and $\beta=1$, yields
\begin{equation}
A^2
=
\frac{\sigma_R^2 L_c}{2\omega_0^\alpha D_0}.
\end{equation}
Equivalently,
\begin{equation}
A
=
\left[
\frac{\sigma_R^2 L_c}{2\omega_0^\alpha D_0}
\right]^{1/2}.
\end{equation}
Here $\sigma_R^2$ is the variance of the projected curvature fluctuations at the microscopic scale, $L_c$ is the correlation length, and $(\omega_0, D_0)$ are reference scales, for example $\omega_0=100\,\mathrm{Hz}$ and $D_0=1\,\mathrm{Gpc}$. Values $A\ll 1$ correspond to weak curvature noise, while $A\sim 1$ would indicate fluctuations large enough to compete with macroscopic curvature contributions over cosmological distances.

\section{Model classification in the $(\alpha,\beta)$ plane}

\begin{figure}[t]
    \centering
    \includegraphics[width=0.9\columnwidth]{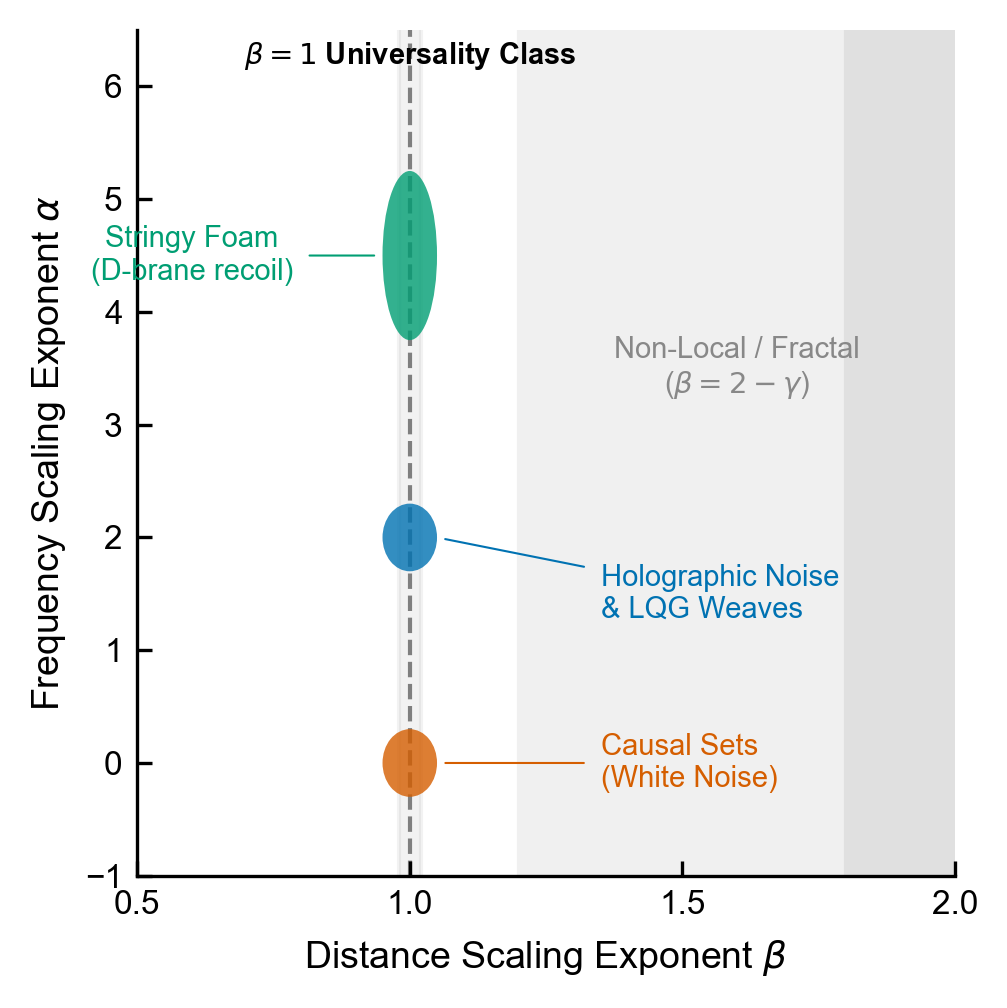}
    \caption{\textbf{Classification of quantum spacetime models in the $(\bm{\alpha}, \bm{\beta})$ scaling plane.} The vertical dashed line at $\beta=1$ represents the \emph{universality class} derived in this work: all local, short-range quantum gravity fluctuations must exhibit linear distance scaling (diffusive random walk). Distinct microscopic theories populate specific "islands" along this line based on their spectral index $\alpha$: Causal Set discreteness ($\alpha \approx 0$), Holographic noise and Loop Quantum Gravity ($\alpha \approx 2$), and energy-dependent String Theory foam ($\alpha \approx 4\text{--}5$). The shaded region to the right ($\beta > 1$) corresponds to non-local or long-range correlations, physically associated with fractal spacetime geometry or coherent astrophysical foregrounds.}
    \label{fig:theory_map}
\end{figure}

Having established that short-range models populate the universal line $\beta=1$, we can use $\alpha$ to distinguish between different microscopic scenarios. Planckian white-noise foam and holographic-type models, in which the microscopic curvature variance is effectively scale-invariant in frequency, correspond to $p=0$ and hence $\alpha=2$. Loop quantum gravity weave states are also expected to produce $\alpha\simeq2$, up to small corrections associated with the discrete area spectrum. Causal set models, where spacetime is realized as a Poisson sprinkling of discrete elements, can produce effective curvature fluctuations that are nearly frequency-independent on relevant scales, leading to $\alpha$ in the range from 0 to 2 depending on the coarse-graining procedure. These expectations are summarized in Fig.~\ref{fig:theory_map}, which maps
representative quantum-spacetime scenarios into distinct “islands” in the
$(\alpha,\beta)$ plane.

String-foam or D-brane recoil models, by contrast, exhibit explicit energy dependence, typically with $n\ge 1$. As we have seen, this leads to $\alpha=2+2n$, i.e., $\alpha\approx4$ for linear recoil and potentially larger values for higher-order couplings. Finally, long-range or nonlocal quantum-spacetime models are characterized by $\beta\neq 1$, with $1<\beta<2$ for power-law tails in the correlator, and can be cleanly separated from local theories on the basis of the distance scaling alone.

\section{Observational signatures and astrophysical foregrounds}

\begin{figure}[t!]
    \centering
    \includegraphics[width=0.9\columnwidth]{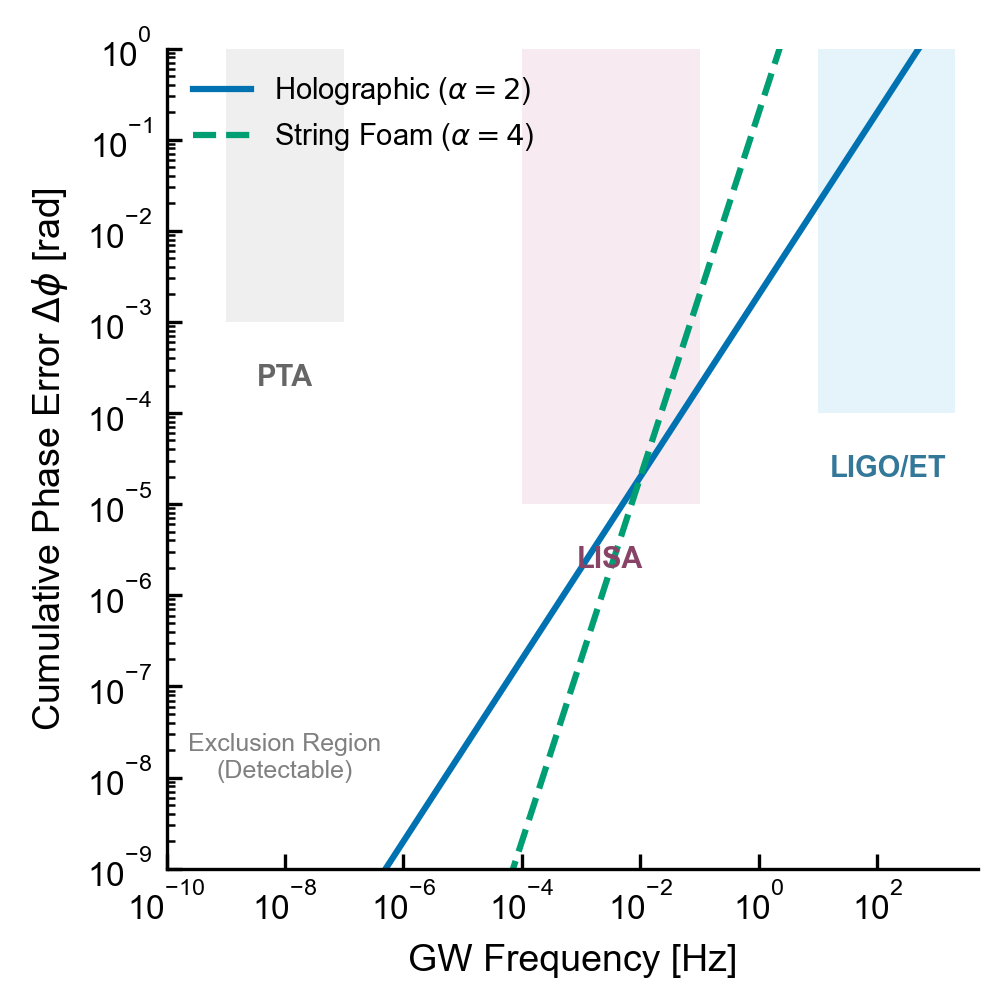}
    \caption{\textbf{Multi-band spectroscopy of quantum spacetime.} The predicted cumulative phase error ($\Delta \phi_{\mathrm{rms}}$) as a function of gravitational-wave frequency for two representative "exotic" quantum gravity scenarios. The curves are normalized to a reference amplitude of $A \sim 10^{-5}$ at LISA frequencies to illustrate the distinct spectral slopes. Shaded rectangles indicate approximate phase-sensitivity floors for Pulsar Timing Arrays (PTA), LISA, and ground-based detectors (LIGO/ET). Stringy Spacetime Foam (dashed green line) predicts a steep frequency dependence ($\alpha \approx 4$, corresponding to $\Delta \phi_{\mathrm{rms}} \propto f^2$), making it most accessible to high-frequency ground-based detectors. Conversely, scale-invariant Holographic noise (solid blue line, $\alpha \approx 2$, corresponding to $\Delta \phi_{\mathrm{rms}} \propto f^1$) is effectively probed by the long baselines of LISA and PTAs. Simultaneous constraints across these bands can break degeneracies and uniquely determine the microphysical exponent $\alpha$.}
    \label{fig:observability}
\end{figure}

The phase diffusion described above manifests observationally as a loss of temporal coherence in the GW signal. In the frequency domain, a nearly monochromatic component (for example, a continuous wave from a rotating neutron star or a quasinormal mode ringdown) is convolved with a Lorentzian kernel of width
\begin{equation}
\Gamma \sim \frac{1}{2} \frac{d}{dt}\langle \Delta\phi^2\rangle,
\end{equation}
resulting in line broadening. For inspiral signals, phase diffusion leads to residual phase noise after subtraction of the best-fit GR template; the variance of these residuals scales with frequency and distance in a way that is distinct from detector noise. Representative predictions for two benchmark quantum-gravity scenarios, together
with the approximate phase-sensitivity floors of PTAs, LISA, and ground-based
detectors, are shown in Fig.~\ref{fig:observability}.

A natural concern is whether classical or astrophysical effects could mimic the same phenomenology. Gravitational lensing by large-scale structure produces Shapiro time delays through integrals of the gravitational potential, but the potential is effectively static on GW timescales. Lensing shifts the mean arrival time but does not produce a random walk in phase, and thus does not generate a diffusion law with $\langle \Delta\phi^2\rangle \propto D$. Dispersion by the interstellar medium produces phase shifts that scale as $\omega^{-1}$ and are negligible at GW frequencies. Dark-matter substructure or inhomogeneous baryonic distributions can modulate the phase through time-varying gravitational potentials, but these correlations extend over kiloparsecs rather than microscopic lengths and tend to produce coherent, rather than diffusive, behavior, more akin to $\beta\approx 2$.

It is therefore difficult for known astrophysical and classical processes to reproduce a universal $\beta=1$ scaling over many independent lines of sight. A robust measurement of $\beta=1$ would strongly favor a local, microscopic origin of the phase noise.

\section{Hierarchical Bayesian inference and constraints}

Individual GW events place relatively weak constraints on $(A,\alpha,\beta)$, since decoherence manifests as a subtle perturbation to the waveform. However, the universality of the effect—arising from propagating through a common class of microscopic fluctuations—allows one to combine many events in a hierarchical Bayesian framework. Let $d_i$ denote the data for event $i$ and $\theta_i$ the usual source parameters. A decoherence-modified waveform can be schematically written as
\begin{equation}
h(t;\theta_i,A,\alpha,\beta) = h_{\rm GR}(t;\theta_i)\,\exp[i\delta\phi_i(t)],
\end{equation}
where $\delta\phi_i(t)$ is a Gaussian random process with variance set by the scaling law (\ref{eq:scaling}). The likelihood for event $i$ is obtained by integrating over realizations of $\delta\phi_i$ and over the source parameters,
\begin{equation}
\mathcal{L}_i(A,\alpha,\beta)
=
\int d\theta_i\,\mathcal{D}\delta\phi_i\,
p(\delta\phi_i|\sigma_i)\,
p(d_i|h_{\rm GR}e^{i\delta\phi_i})\,
p(\theta_i),
\end{equation}
where $\sigma_i^2 = A^2(\omega_i/\omega_0)^{\alpha}(D_i/D_0)^{\beta}$ encodes the variance for that event.

The joint posterior for $(A,\alpha,\beta)$ given $N$ events is then
\begin{equation}
p(A,\alpha,\beta|\{d_i\})
\propto
p(A,\alpha,\beta)
\prod_{i=1}^N \mathcal{L}_i(A,\alpha,\beta),
\end{equation}
with $p(A,\alpha,\beta)$ a prior informed by theoretical expectations. In practice, the effect of decoherence can be parameterized directly at the level of phase residuals or ringdown linewidths, leading to tractable likelihoods. The combination of ground-based detectors (sensitive to high-frequency, high-$\alpha$ effects), LISA (sensitive to long baselines), and PTAs (probing ultra-long-baseline, low-frequency signals) provides complementary leverage on the parameter space.

A null result, i.e., the absence of detectable phase diffusion down to $A\lesssim 10^{-5}$ for $\beta=1$, would translate into upper bounds on $\sigma_R^2 L_c$ and therefore into constraints on the effective correlation length $L_c$ and microscopic scales of quantum geometry. In particular, it would allow one to rule out classes of “exotic” models with $\ell_{\rm eff}\gg \ell_P$ or with macroscopic $L_c$.

\section{Conclusion}

We have presented a gauge-invariant, adiabatic framework for calculating the decoherence of gravitational waves in a quantum spacetime. By identifying the phase variance as the unique leading-order observable in the deep adiabatic regime, we derived a universality theorem: all local, short-range quantum gravity models must exhibit linear distance scaling ($\beta=1$) for cumulative GW phase diffusion. This universality reflects a geometric version of the Central Limit Theorem applied to curvature fluctuations.

The result simplifies the phenomenological landscape and allows us to categorize theories based on the spectral exponent $\alpha$. Models such as Planckian foam, holographic noise, causal sets, loop quantum gravity, and string foam occupy distinct regions in the $(\alpha,\beta)$ plane, offering a clear avenue for falsifiability. Although standard Planck-scale fluctuations are unlikely to be directly observable, the scaling laws derived here turn the Universe into a precise filter: even a null detection of decoherence carries significant information about the microscopic structure of spacetime and the range of quantum-gravity correlations.

As GW observations accumulate across multiple frequency bands and baselines, the framework developed in this work can be used to either detect the spectral fingerprint of quantum geometry or to confine it to ever smaller scales, thereby placing increasingly stringent constraints on the granularity and locality of spacetime.


\end{document}